# Transient Work Function Gating:
# A New Photoemission Regime

S. Carbajo[1,§]
[1]*SLAC National Accelerator Laboratory and Stanford University, 2575 Sand Hill Road, Menlo Park, CA 94025, USA*

§corresponding author: scarbajo@stanford.edu

(Dated: 2 June 2020)

We present the theoretical basis for a new photoemission regime, transient work function gating (TWFG), that temporally and energetically gates photoemission and produces near-threshold photoelectrons with thermally limited emittance, percent-level quantum efficiency, and control over temporal coherence. The technique consists of actively gating the work function of a generalized photocathode using non-ionizing long-wavelength optical field to produce an adiabatic modulation of the carrier density at their surface. We examine TWFG as a means to circumvent the long-standing trade-off between low emittance and high quantum efficiency, untethered to particle source or photocathode specifics. TWFG promises new opportunities in photoemission physics for next generation electron and accelerator-based x-ray photon sources.

## 1. INTRODUCTION

Today's electron sources underpin transformational scientific instrumentation, most notably colliders[1], and ultrafast electron[2] and X-ray free electron lasers (XFEL)[3]. The latter have enabled myriad scientific discoveries in physical chemistry[4–6] and biology[7–9] with insurmountable societal impact. Their advent has revolutionized the paradigm for atomically resolving ultrafast dynamics in a broad spectrum of scientific research[10]. But increasing these scientific instruments' capabilities is paramount to advancing exploration of fundamental atomic and molecular dynamics in uncharted waters. The operational capacity of these scientific instruments is largely dictated by the electron beam brightness and emittance, and in particular from the photoemission processes that generates and accelerates the initial electron beam in vacuum, namely the photoinjector.

Traditional photoinjectors are used to produce photoelectrons using first laser pulses with photon energies above the photocathode work function, typically in the ultraviolet (UV) and up to the visible range. The photoelectrons are then quickly accelerated by radiofrequency (RF) fields to generate electron beams in vacuum. There are two established strategies to increase the photoinjector brightness: increasing the gun accelerating gradient to increase the extractable charge density and reducing the transverse electron energy spread[11,12]. On the latter, the most salient quantity to characterize the transverse energy spread is the intrinsic—also named thermal—emittance ($\epsilon_{i,x}$) of the photocathode. The intrinsic emittance imposes a lower limit for the normalized emittance that can be generated by a photoinjector, and by association, the upper limit for the brightness of any source that cascades from the photoinjector, such as storage rings, linear accelerators and X-ray free electron lasers (XFEL). There are multiple challenges associated with achieving bright electrons already from the photoemission process itself. In general, the intrinsic emittance—or mean transverse energy (MTE) of an electron beam—is proportional to the difference between photocathode work-function ($\phi_w$) and photon energy ($h\nu$) of the laser used to produce photoelectrons. In the current photoemission paradigm, there is a competition between emittance and charge yield. Minimizing ($h\nu - \phi_w$) to near-threshold emission provides arbitrarily low MTE values per the Fermi-Dirac (F-D) distribution. But near-threshold emission also results in low quantum efficiencies (QE), a quantity that represents the number of electrons that exit the photocathode into vacuum per absorbed photon. Practically, low QEs demand proportionately high laser powers for the any given charge yield, which can cause detrimental heating and nonlinear effects[13]. These in turn may cause localized heating and multi-photon absorption, which are bound to spread the F-D distribution and thus the beam MTE and emittance.

The phase-space distribution of an electron beam from a photoinjector is also contingent on the spatio-temporal distribution of the photoexcitation laser. Uniformly-filled ellipsoidal charge distributions are known to best mitigate space-charge induced phase-space dilution because they produce space-charge fields with linear dependence on position within the distribution[14,15]. They are also less prone to halo formation, which is particularly attractive for kW-level XFELs[16]. But temporally shaping of photoexcitation lasers suited for high average power XFEL operation, particularly in the ultraviolet (UV) as it is the case in conventional photocathode materials, is also a complex task undertaken under various linear and nonlinear optics efforts[17].

## 2. NEW PHOTOEMISSION REGIME

The main premise of transient work function gating (TWFG) is that a non-ionizing strong field can energetically and temporally control the carrier dynamics and adiabatically modulate the energy bands of a generalized photocathode. This process can be intuitively understood as a short-lived abrupt Schottky barrier that temporarily reduces the photocathode work-function before it is restored to its original state of equilibrium. Gated photoemission in this regime could be implemented, for example, with a monochromatic laser pulse with photon energy below the photoemission threshold in combination with synchronous far infrared (FIR) or terahertz (THz) E-field transient as a gate, such that the monochromatic pulse will only cause photoemission during the transient state. The most prominent feature of this technique is that ultralow intrinsic emittances at moderate to high QEs can be achieved through the interplay between the Schottky barrier strength and the photon energy of the photo-excitation laser. The gating fields will lead to oscillatory non-thermal electron transitions and modify the Fermi energy ($E_F$). This disturbance will cause the energy bands to change alongside $E_F$, thereby enabling near threshold emission (i.e. with thermally limited emittance) without paying the customary QE penalty.

Today, efficient long-wavelength nonlinear optical generation techniques[18–20] are sufficiently mature to render accelerator concepts and demonstration in lieu of conventional radio-frequency (RF) accelerators[21–23]. In this frequency range, the field emission threshold for surface electric field and breakdown limit increases beyond the GV/m for THz and well into the TV/m for FIR depending on the peak and average power loads[24]. TWFG capitalizes on this particular advantage to create abrupt Schottky barriers without inducing material breakdown. In the next two sections, we present the theoretical basis for TWFG and a case study relevant for linear accelerators and XFELs as an exercise of future experimental applicability.

## 3. BASICS OF TRANSIENT WORKFUNCTION GATING

We use the three-step model for photoemission[25] as the baseline for the TWFG model. The three-step model is physical representation that accurately describes photoemission in both metallic and semiconductor photocathodes[26] where (i) a photon with energy $h\upsilon$ is absorbed within a few-atomic layers of the photocathode, (ii) a photoelectron scatters and migrates to the surface, and (iii) finally the electron escapes above the barrier to vacuum from the Schottky effect and change in angle. In this formalism, we consider the photocathode work function $\phi_w$ to be an intrinsic yet adaptable property. In this section, we extend the three-step model to include TWFG, describe the ultrafast modification of the Schottky barrier and the abrupt change in electron angle across the photocathode-vacuum interface, and derive expressions for the intrinsic emittance and quantum efficiency.

Let us begin with the F-D distribution in the photocathode. The electron density of occupied states and the electric potentials experienced by a single electron immediately outside the cathode are shown in Figure 1. The Schottky barrier is the sum of the image charge field, naturally arising from the occupied density of states, and an externally applied field ($F_a$). The Schottky work function, $\phi_{Schottky}$, is defined as the peak value of the Schottky barrier and represents the photoemission threshold potential that is located a few nanometers away from the photocathode-vacuum interface. A non-zero $F_a$ field will reduce the photoemission threshold potential and shift the zero-field vacuum state downward, as shown in green, thus increasing the quantum yield for a given photon energy. Thus, the effective work function, $\phi_{eff}$, in the presence of the Schottky barrier is

$$\phi_{eff} = \phi_w - \phi_{Schottky} \qquad \text{(Eq. 1)}$$

and for a static applied field, $F_a$, where the material response is much faster than the field period

$$\phi_{eff} = \phi_w - e\sqrt{\frac{eF_a}{4\pi\varepsilon_0}} = \phi_w - 0.037947\sqrt{F_a\left(\frac{\text{MV}}{\text{m}}\right)} \text{ [eV]} \qquad \text{(Eq. 2)}$$

where $e$ is the elementary charge and $\varepsilon_0$ is the vacuum permittivity. The static $F_a$ field approximation in (Eq. 2) is generally applicable regardless of the frequency so long as the material response time is faster than the field. Let us provide a few examples in current technologies to illustrate the concept of effective work function. Note that direct current (DC) guns, for instance, can only allow relatively low static accelerating gradients because of breakdown limits. In the case of high-gradient electron guns, such as superconducting radio frequency (SRF) guns, the applied field is in the order of few tens of MV/m at most and durations of several tens of ns or longer, that is, orders of magnitude longer than the material response times, thus practically static in comparison to photoemission timescales where $\phi_{Schottky}$ is a long-lived barrier at ~100 meV energy. To take Cu as an example, $\phi_{eff}$ is reduced from 4.31 eV to 4.14 eV with a 20 MV/m field, which is why many photoinjectors today operate at ~266 nm wavelength photoexcitation (4.66 eV), corresponding to the third-harmonic of Ti:Sapphire laser systems. But substantially higher field strengths are accessible with long-wavelength optical fields in the FIR and up to the THz region. To follow on the previous simple example, at 1 GV/m peak field, which is well below material breakdown for FIR pulses, yields $\phi_{eff} = 3.1$ eV, now a significant Schottky barrier reduction.

We will further describe the ability to generate an abrupt work function modulation from long-wavelength gating fields with the following added considerations to the three-step model:

i. The material response is no longer negligibly small compared to the field period and thus must be considered not only in computing $\phi_{eff}$ but also in the TWFG model

ii. The field emission and breakdown limit can be extended approximately to the single-digit GV/m for THz frequencies and up to the TV/m range in the case of FIR frequencies, depending on the specific material, pulse duration, and average power conditions[27–29], which can reduce $\phi_{eff}$ by > 1 eV and enable visible range photoexcitation in conventional photocathodes.

iii. The gating fields create a transient state of the intrinsic optical and electronic properties of the material arising from a field-dependent adiabatic carrier concentration modulation.

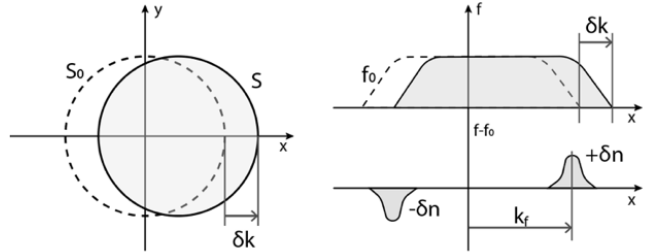

Figure 1: effect of gating field on momentum space and Fermi distribution displacement

Let us examine the long-wavelength field induced carrier dynamics. We denote $E_F$ as the highest energy occupied by electrons. Figure 1 shows an initial Fermi surface in thermodynamic equilibrium (S$_0$). When an external (gating) electric field is applied, $E_{gate}$, with its vector aligned along $x$, the original Fermi surface S$_0$ will be displaced in momentum k-space from its original state (f$_0$) by $\delta k$ along $x$ to a modified transient Fermi surface (S), in which the state density will change, and so will $E_F$. That is, when an external field is present, the Fermi distribution ($f$) shifts to higher occupation states until it vanishes, the point at which the Fermi surface relaxes back to the equilibrium distribution by means of electron scattering from the temporarily occupied states ($+\delta n$) to the unoccupied states ($-\delta n$). For a finite $E_{gate}$, this shift ($\delta n$) will be proportional to the field strength. In essence, the new Fermi distribution exists only while the external field is on. Following Drude Theory[30], we calculate the momentum displacement as

$$\delta k = -\frac{eE_{gate}\tau}{\hbar} \quad (Eq. 3)$$

where $\tau$ is the relaxation time constant of the material. Note that the typical velocity of electrons is very large, and that for a metal scattering lengths are about 100 Å at room temperature and ~1 mm at low temperature.

In energy space, this Fermi shift equates to

$$\delta E = -\frac{(eE_{gate}\tau)^2}{2\,m_e} \tag{Eq. 4}$$

Now that we have defined the momentum displacement of the Fermi distribution, let us focus on the application of this shift to near-threshold photoemission (NTP), that is, where $h\nu - \phi_{eff} \lesssim 300$ meV. Extensive studies of the three-step model[25,31–35] demonstrate that in this regime the photon absorption conserves energy but not momentum. Here, electrons can be considered as non-interacting fermions that occupy states as given by the gated F-D distribution and the QE can be expressed as[33]

$$QE(\omega,\delta E) = \frac{1-R(\omega)}{1+\lambda_{opt}(\omega)/\lambda_{e-e}(\omega)} \frac{(E_F+\delta E+h\nu)}{2h\nu}\left[1-\sqrt{\frac{E_F+\delta E+\phi_{eff}}{E_F+\delta E+h\nu}}\right]^2 \tag{Eq. 5}$$

where $R(\omega)$ is the photocathode optical reflectivity as a function of optical frequency, $\omega$; $\lambda_{opt}(\omega)$ is the photon absorption length; and $\lambda_{e-e}(\omega)$ is the electron-electron mean-free path. In NTP, because of a relatively small energy range, $\lambda_{e-e}(\omega)$ is nearly constant and its average value, $\hat{\lambda}_{e-e}$, can be used. Likewise, because of the narrow bandwidth of monochromatic photoexcitation pulses centered at optical frequency $\omega_0$, $R(\omega)$ and $\lambda_{opt}(\omega)$ can also be reduced to their nominal values, $R(\omega_0)$ and $\lambda_{opt}(\omega_0)$, respectively. Eq. 5 is a valid representation of the QE both under ungated (normal) and gated conditions. Note that we do not apply a Taylor series reduction approximation of Eq. 5, as it is customary in estimating QE, since we must solve for the full equation in NTP.

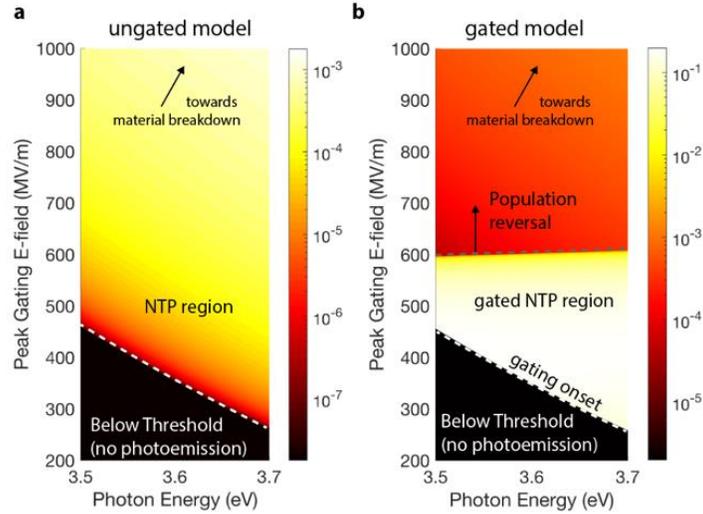

Figure 2: QE projections for a Cu photocathode under a) normal and b) three-step photoemission model that includes transient gating as a function of peak gating field and photoexcitation energy. Colormaps have different scales.

Figure 2 shows two QE maps as a function of peak gating field and photoexcitation photon energy for the gated and ungated QE model based on Eq. 4 and Eq. 5 using a Cu photocathode as an example. The effects of transient work-function gating are palpable by a two order of magnitude increase in achievable QEs. In Figure 2.a, the ungated QE reaches up to $1.8 \cdot 10^{-3}$ at the highest photon energy and highest peak gating field, as expected. This is because the traditional model only considers the difference between photoexcitation energy and a static field. By using Eq. 5, which incorporates transient Fermi displacement, we see the interplay between a few interesting mechanisms in Figure 2.b. Just like for the ungated case, there's a moderate E-field level below which no photoemission is possible, at least in the linear absorption regime. The gating onset becomes apparent when sufficiently strong fields are present—in this example above ~260 MV/m—where the Fermi displacement contribution becomes strong enough to access the NTP region. Foreseeably, the onset takes place approximately when $h\nu \approx \phi_{eff}$. Above the onset, the gated QE asymptotically approaches its maximum level, in this case 0.19, although it is consistently >10% for peak gating fields ranging from about 500 to 600 MV/m. More generally, the gated QE is maximized in the NTP region when the net Fermi displacement and the photon energy are comparable, that is, when $|E_F + \delta E| \approx h\nu$.

The interplay between scattering and optical effects yield a wider region of high gated QEs as the photoexcitation energy increases. Continuing on above the gated NTP region, in this case above ~600 MV/m, the fields can be detrimental to effective extraction of charge by creating a sufficient population reversal that shields electrons from exiting the material. This carrier population reversal causes the QEs to be slightly lower compared than those projected in the ungated model. Continually increasing the field strength towards a significant fraction of a TV/m or above, depending on the specific mechanical and optoelectronic properties of the material, sustained strong gating fields could begin to cause permanent material breakdown.

It is important reiterating that the %-level QEs are only valid in *gated* NTP. That is, if the photon energy were to be extended to levels significantly above threshold, the gated QE would then equate the ungated QE, since at that point $\delta E$ is small compared to $|h\nu - \phi_{eff}|$. It is also worth noting that the optimal cooperation between the Fermi displacement and the QE depends on the intrinsic material optoelectronic properties and conditions, including temperature and lattice structure, where applicable. But the primary conclusion from this example is that the QE levels for standard materials like Cu are dramatically increased by remaining in the gated NTP region depicted in Figure 2.b.

Let us now continue on with the benefits of this new photoemission process by focusing on emittance. The intrinsic emittance, $\epsilon_{i,x}$, is by definition a measure of the transverse momentum of the electrons that are emitted, which is related to the mean transverse energy (MTE) by the following expression:

$$\epsilon_{i,x} = \sigma_x \sqrt{\frac{MTE}{mc^2}} \tag{Eq. 6}$$

where $\sigma_x$ is the initial rms size of the beam along a transverse Cartesian coordinate $x$ and $mc^2$ is the electron rest energy. The MTE is analogous to the temperature of the photoemitted electrons. For photon energies much larger than the work function, $h\nu \gg \phi_{eff}$, we can approximate $MTE = (h\nu - \phi_w)/3$ because of a relatively constant density of states in the portion of the conduction band accessible by a photon of a given energy[36]. However, in the case of NTP, this approximation is no longer valid because photoemission can still occur from below-Fermi occupation states or from impurity states within the band gap in the case of semiconductors[25]. One way of reducing emission from these states is to decrease the photon energy and cathode temperature. By doing so, the MTE and QE resembles that of what is expected from the tail of the F-D distribution and the NTP MTE can be expressed as[37]

$$\text{MTE} = kT \left( \text{Li}_3 \left( -\exp\left(\frac{h\nu - \phi_{eff}}{kT}\right) \right) \Big/ \text{Li}_2 \left( -\exp\left(\frac{h\nu - \phi_{eff}}{kT}\right) \right) \right) \tag{Eq. 7}$$

where $\text{Li}_s$ is the polylogarithm function of order s, $T$ is the photocathode temperature, and $k$ is the Boltzmann constant. It is important to note that as $|h\nu - \phi| \to 0$, that is, very close to the photoemission threshold, MTE = $kT$. That is, if electrons are shifted from the Fermi energy towards the minimum possible emission energy by, for instance, lowering a small band gap photocathode temperature, and photoemission occurs nearly at threshold, the intrinsic emittance could be reduced to near the F-D distribution limit (single-digit meV) since the emitted electrons have less available transverse momentum. This is showcased in Figure 3 for two NTP photoexcitation energies where thermally limited emittances are possible at above-10% QEs. It is important to note once again that without the cooperative interplay between the transient Fermi displacement ($\delta E$) caused by the gate and photoexcitation in NTP this would not be possible. This is precisely why traditional (i.e. ungated) NTP operation yields very low charges with QE < 10-5, as shown in the ungated (dashed) QE projection.

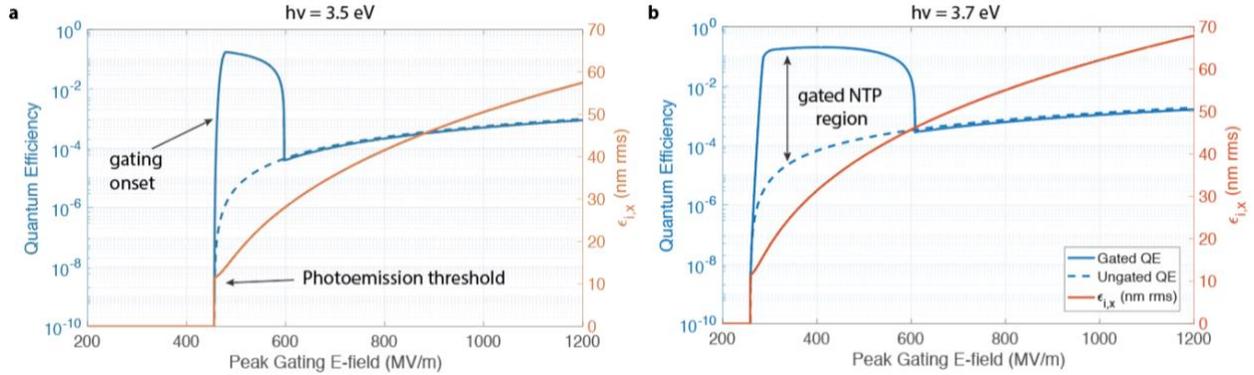

Figure 3: ungated (dashed) and gated (solid) QEs for two NTP energies a) $h\nu = 3.5$ and b) $3.7$ eV in Cu (T=70K) and the corresponding intrinsic emittance as a function of peak gating field ($\sigma_x = 100$ µm)

### 4. CASE STUDY: HIGH-BRIGHTNESS GUNS FOR LINEAR ACCELERATORS AND XFELs

The primary objective of this section is to employ the framework described above to exercise a few quantifiableF metrics for its application in high-brightness electron guns for linear accelerators and XFELs, that is, at practically high charge levels.

Let us focus on ultrafast work-function gating for generating electron beams—single bunch and trains—with ultralow emittance. Here, we employ a regular Cu photocathode cooled at cryogenic temperatures. We consider the material response and employ higher frequency higher strength gating fields with two representative waveform shapes as possible gating field candidates: single-cycle (Figure 4.a) and multi-cycle (Figure 4.b). All other relevant simulation parameters are listed in Table 1.

We present the single-cycle and multi-cycle fields because they represent two interesting emission scenarios. The single-cycle regime resembles a more traditional electron bunch generation scheme, where only one electron bunch will be emitted per pulse. In the multi-cycle scenario, a train of electron bunches spaced by the period of the gating field will be emitted even when using a single photoexcitation laser pulse, thereby enabling the generation of longitudinally bunched electron beams. This temporally shaped emission of the electron bunches could be ideally suited for serial ultrafast electron diffraction (UED), where multiple scattering events could be acquired within one exposure as an equivalent to serial femtosecond crystallography. It could also be exploited for X-ray generation via inverse Compton scattering or bunched to increase a single peak brightness for colliders, for instance.

Figure 4-a and -b show the longitudinal charge distribution emitting off the surface of the photocathode from of a single-cycle and multi-cycle gating waveform with fixed peak amplitude, respectively, for photoexcitation energies from 3.5 to 3.7 eV. All figures show the time-evolution of projected emittances remaining at the few 10s of nm rms with enhanced QEs ranging from 2-11% depending on the exact photoexcitation laser photon energy. Both photoexcitation and gating pulses are assumed to be time-overlapped, where the photoexcitation pulse width is arbitrarily longer than the gate pulse width.

There are a few noteworthy points specific to this example. First, ultralow intrinsic emittances with %-level QEs would not be possible without the transient gate fields, which is why pC charges are generated with few-nJ photoexcitation energies at NTP. The photoexcitation wavelengths are now easier to produce and handle (~350 nm) than the deeper UV that Cu would require to use without gating (< 275 nm). For example, if we were to consider a lower work-function photocathode such as $Cs_2Te$ ($\phi_w = 3.8$ eV), the equivalent photoexcitation wavelengths with transient gates would actually shift to the blue-green spectrum (440-480 nm). Longitudinally shaped charge distributions are accessible because photoemission is longitudinally shaped by the gating fields (and not intensity), thereby bringing new alternative solutions to longitudinal shaping of electron bunches directly off the photocathode complementary to UV[38] or laser heater shaping[39] techniques. Note that generating single- or multi-cycle gating fields with tunable waveforms in the TWFG regime is currently possible via optical rectification and waveform shaping in nonlinear[40] and quasi-phase matched[19] crystals. As a result, many other longitudinal photoemission distributions are possible beyond these two examples.

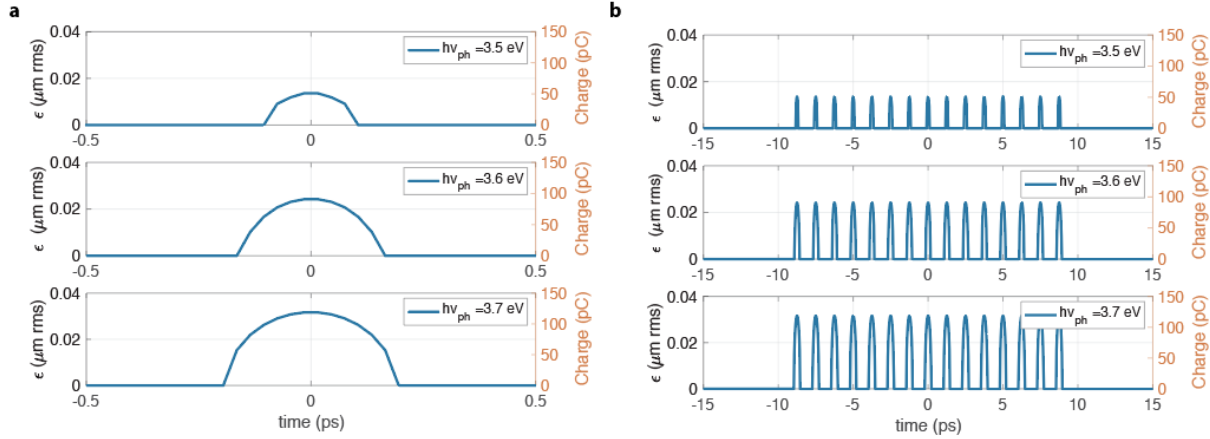

Figure 4: time-evolution of gated emittances and charge from Cu (T=70 K) using single-cycle (a) and multi-cycle (b) fields for various photoexcitation energies.

## 5. FINAL REMARKS

The potential impact of a new photoemission regime as profound as it is foundational. First, emitting thermally limited electron bunches at practically high QEs and charges cannot be achieved under the current photoemission paradigm and circumventing this trade-off could transform the production of electron beams with increased average flux and peak brightness as well as scaling the brightness of accelerator-based light sources, especially XFELs.

To take the LCLS-II gun as an example, we estimate that this new technique could provide normalized emittances as low as ~0.1-0.15 µm rms with 200-300 pC charges, which is orders of magnitude lower than currently projected. Practical operation of XFELs with such low emittances would enable proportionately higher extension of their brightness to even harder X-rays or a significant reduction of undulator lengths for the same X-ray photon energy. XFELs operating in the harder X-ray regime are technically demanding and to achieve laser saturation the undulator must be 100s of m long. Reducing emittance makes the FEL process more efficient because it achieves gain saturation quicker, thereby requiring a shorter undulator and reducing cost. Temporally controlling photoemission can also enable much greater operational capacity of XFELs by providing orders of magnitude higher temporal coherence approaching the Fourier-transform limit for high-resolution spectroscopies, such as (resonant) inelastic X-ray scattering, or by providing high brightness multi-beam mode operation for multi-color X-ray experiments at harder X-rays or for self-seeded operation with improved coherence and monochromaticity.

**ACKNOWLEDGEMENTS**
This work was supported in part by the U.S. Department of Energy and Laboratory Directed Research and Development program at SLAC National Accelerator Laboratory, under Contract No. DEAC02-76SF00515.

**DATA AVAILABILITY**
The data that supports the findings of this study are available within the article.

**Methods**

Table 1: Simulation parameters

| Photocathode material | | Cu |
|---|---|---|
| Temperature (K) | $T$ | 70 |
| Fermi Energy (eV) | $E_F$ | 7 |
| Workfunction (eV) | $\phi_w$ | 4.31 |
| electron mean free path (nm) | $\hat{\lambda}_{e-e}$ | 2.2 |
| Photon mean free path (nm) | $\lambda_{opt}$ | 28.6 |
| Photoex. energy (eV) | $\hbar\omega_0$ | 3.5-3.7 |
| Photoex. pulse energy (nJ) | $E_{h\nu}$ | 5 |
| Transverse spot size (μm) | $\sigma_x$ | 80 |
| Gate peak field (GV/m) | $E_{gate}$ | 0.5 |